\documentclass[twocolumn,a4paper,superscriptaddress,floatfix]{revtex4}
\usepackage[english]{babel}
\usepackage[utf8]{inputenc}
\usepackage{amsmath}
\usepackage{graphicx,epstopdf}
\usepackage[colorinlistoftodos]{todonotes}
\usepackage{amssymb,epsfig,color,textcase}
\usepackage{hyperref}
\usepackage{doi}

\begin{document}

\title{Exchange constants of the Heisenberg model 
in the plane-wave based methods using the Green's function approach}

\author{Dm.M.~Korotin}
\affiliation{Institute of Metal Physics, S.Kovalevskoy St. 18, 620990 Yekaterinburg, Russia}
\email{dmitry@korotin.name}

\author{V.V.~Mazurenko}
\affiliation{Department of theoretical physics and applied mathematics, Ural Federal University, Mira St. 19, 620002 Yekaterinburg, Russia}

\author{V.I.~Anisimov}
\affiliation{Institute of Metal Physics, S.Kovalevskoy St. 18, 620990 Yekaterinburg, Russia}
\affiliation{Department of theoretical physics and applied mathematics, Ural Federal University, Mira St. 19, 620002 Yekaterinburg, Russia}

\author{S.V.~Streltsov}
\affiliation{Institute of Metal Physics, S.Kovalevskoy St. 18, 620990 Yekaterinburg, Russia}
\affiliation{Department of theoretical physics and applied mathematics, Ural Federal University, Mira St. 19, 620002 Yekaterinburg, Russia}
\email{streltsov@imp.uran.ru}

\date{\today}

\begin{abstract}
An approach to compute exchange parameters of the Heisenberg model in plane-wave based methods is presented. This calculation scheme is based on the Green's function method and Wannier function projection technique.  It was implemented in the framework of the pseudopotential method and tested on such materials as NiO, FeO, Li$_2$MnO$_3$, and KCuF$_3$. The obtained exchange constants are in a good agreement with both the total energy calculations and experimental estimations for NiO and KCuF$_3$. In the case of FeO our calculations explain the pressure dependence of the N\'{e}el temperature. Li$_2$MnO$_3$ turns out to be a Slater insulator with antiferromagnetic nearest neighbor exchange defined by the spin splitting. The proposed approach provides a unique way to analyze magnetic interactions, since it allows one to calculate orbital contributions to the total exchange coupling and study the mechanism of the exchange coupling. 
\end{abstract}

\maketitle

\section{Introduction}
Magnetic interactions in modern materials are in the focus of the theoretical and experimental investigations.
Depending on the nature and localization of the magnetic moments one can use different model Hamiltonians to describe the magnetic properties of the system.
In case of the localized magnetic moments the spin Hamiltonian approach based on the solution of the Heisenberg model can be uses. The corresponding Heisenberg Hamiltonian has the form 
\begin{equation}
\label{Heis_q}
H_{Heis} = \sum_{\langle ij \rangle} J_{ij} \hat {\bf S}_i \hat {\bf S}_j,
\end{equation}
where $J_{ij}$ is the isotropic exchange interaction parameters.
One can also use different extensions of the Heisenberg model taking into account symmetric
and antisymmetric parts of the anisotropic exchange coupling~\cite{Eremin,blundell-magnetism,Khomskii2014}.
Within the spin Hamiltonian approach the problem of realistic description of the magnetic properties is reduced to the problem of unambiguous determination of the exchange interactions by taking electronic structure and chemical bonding into account. It  can be done on different levels and by using different means.

One of the most popular approaches for {\it ab initio} investigation
of solids is density functional theory (DFT). There are a few
methods to estimate exchange constants $J_{ij}$ within DFT, i.e.
 to map the results of the DFT calculations onto the Heisenberg
model. 

The most direct, and popular way to calculate $J_{ij}$ is to calculate 
the total energies of the $N+1$ magnetic configurations, where
$N$ is the number of different exchange constants~\cite{Noodleman1981,Martin-book,Tsirlin2014}.
Despite the robustness of this approach, it has several serious drawbacks: 
(1)~a number of different magnetic configurations
have to be calculated for complicated systems; (2)~all configurations must use the same magnetic moments 
(important for the materials close to itinerant regime); and 
(3)~the result is purely a number, which is hard to analyze, i.e.
understanding which orbitals contribute the most and what 
mechanism of exchange coupling (direct exchange, super-exchange,
double exchange etc.) is present.

To overcome these shortcomings the Green's function method~\cite{LEIP1,Liechtenstein1987,Katsnelson2000} can be utilized. 
Using DFT and Heisenberg models, it produces analytical expressions for the changes in the total energy with respect to small spin rotations. 
This approach allows one not only to obtain all the exchange constants from the calculation of a single magnetic configuration, but also
to find contributions to the total exchange coupling coming from different orbitals (i.e., e.g.
$J_{xy/xy}$, $J_{xy/x^2-y^2}$ etc.).
Moreover, this method can easily be generalized to
calculate the anisotropic part of the exchange 
Hamiltonian~\cite{Mazurenko2005}.

Previously, the Green's function approach was formulated for localized orbitals
methods, e.g. linear muffin-tin
orbitals (LMTO) method~\cite{Andersen1984} or linear combination
of atomic orbitals (LCAO)~\cite{Bloch1928,Harrison1999}. 
However, modern high-precision schemes of band structure 
calculations are mostly based on the methods, which use a plane-wave-type basis. They are the full-potential (linearized) augmented plane-wave (L)APW~\cite{Singh-book} and pseudopotential~\cite{Martin-book} methods. As a result,
a straightforward realization of the Green's function
method becomes impossible within plane-wave approaches and all its
advantages cannot be used in the modern {\em ab initio} DFT
codes without direct definition of a localized basis set.

In the present paper we show how the Green's function
approach can be adapted for the plane-wave based 
methods using the Wannier functions formalism.
We implemented this calculation scheme in the
pseudopotential Quantum-ESPRESSO code~\cite{Giannozzi2009} and
report the results concerning the magnetic interactions in
NiO, FeO, Li$_2$MnO$_3$, and KCuF$_3$.

\section{Method \label{Method}}
Following Ref.~\cite{Liechtenstein1987}, we used classical analogue 
of Eq.~\eqref{Heis_q} with spins substituted by the unit vectors 
${\bf e}_i$ pointing in the  direction of the $i$th site magnetization:
\begin{equation}
\label{LP}
H = \sum_{\langle ij\rangle} J_{ij} {\bf e}_i {\bf e}_j.
\end{equation}
The value of the exchange constants for the conventional classical 
Heisenberg model (with spins, not unit vectors) can be obtained
with a proper renormalization.

The power of the Green's function method is in the application
of the local force theorem (see e.g. Ref.~\cite{Methfessel1982}).
When the spins experience rotations over a small angle $\delta \phi$, 
the resulting change to the spin density in the DFT can have the local force theorem applied~\cite{Liechtenstein1987}. 
This can only be done if the Hamiltonian of the system is defined in a localized
orbitals basis set (otherwise it is not clear what parts of the Hamiltonian
have to be rotated). The result of the rotation is compared with a 
similar procedure performed for the spin Hamiltonian~\eqref{LP},
which allows us to derive an analytical expression for the exchange
integrals ~\eqref{JLEIP}. 
The major difficulty in the application of this approach to the modern 
plane-wave based calculation schemes is the absence of the
localized basis set in these methods. We propose to use the Wannier
functions (WF) projection procedure to avoid this restriction
and show its realization for the pseudopotential method.

It is important to note that the Heisenberg model is defined
for localized spin moments. Therefore the basis set with the
most localized orbitals is the best for a mapping of the DFT results  on the Heisenberg model. Hence the maximally localized Wannier functions~\cite{Marzari1997} represent most natural choice for such a mapping. Technically the localization degree and the symmetry of such wavefunctions can be controlled in the projection procedure. One of the most widespread procedures is an enforcement of maximum localization of WF~\cite{Souza2001}. The second~\cite{PhysRevB.71.125119} is a constraint for the WF symmetry to be the same as the symmetry of pure atomic $d$-orbitals. 
In the present paper the second type of projection procedure is used. 

The WFs were generated as projections 
of the pseudoatomic orbitals 
$|\phi_{n{\bf k}}\rangle = \sum_{\bf T} e^{i{\bf kT}}|\phi_n^{\bf T}\rangle $ onto 
a subspace of the Bloch functions $|\Psi_{\mu{\bf k}}\rangle$ 
(the detailed description of WFs construction procedure within 
pseudopotential method is given in Ref.~\cite{Korotin2008a}):
\begin{equation}
|W^{\bf T}_n\rangle = \frac{1}{\sqrt{N_{\bf{k}}}} \sum_{\bf k} |W_{n{\bf k}}\rangle e^{-i{\bf kT}},
\end{equation}
where
\begin{equation}
\label{Wannier:proj}
|W_{n{\bf k}}\rangle \equiv \sum_{\mu=N_1}^{N_2} |\Psi_{\mu{\bf k}}\rangle\langle \Psi_{\mu{\bf k}} | \phi_{n{\bf k}} \rangle.
\end{equation}
Here ${\bf T}$ is the lattice translation vector. The resulting WFs have the symmetry 
of the atomic orbitals $\phi_n$ and describe the electronic states that
form energy bands numbered from $N_1$ to $N_2$. 

The matrix elements of the one-electron Hamiltonian in the reciprocal space are defined as:
\begin{equation}
\label{Wannier:Ham-k}
H^{WF}_{nm,\sigma}({\bf k}) = \langle W_{n{\bf k}} | \left (  \sum \limits_{\mu=N_1}^{N_2} |\Psi_{\mu{\bf k}} \rangle \varepsilon_\mu^{\sigma}({\bf k})\langle \Psi_{\mu{\bf k}} | \right ) | W_{m{\bf k}} \rangle ,
\end{equation}
where $\varepsilon_\mu^{\sigma}({\bf k})$ is the eigenvalue of the one-electron Hamiltonian for band $\mu$ and spin $\sigma$.

Such a Hamiltonian matrix is produced as a result of the WF 
projection procedure at the end of the self-consistent cycle 
in the spin-polarized DFT or DFT+U calculations. 

This matrix in the $H^{WF}_{mm',ij,\sigma}$ form
(where $m$ and $m'$ numerate orbitals on $i$th and
$j$th sites, respectively) can be used for the inter-sites Green's function calculation at every ${\bf k}$-point in reciprocal space:
\begin{equation}
\label{Green}
G^{mm'}_{ij,\sigma}(\varepsilon,{\bf k}) = {(\varepsilon+E_F-H^{WF}_{mm',ij,\sigma}({\bf k}))}^{-1},
\end{equation}
where $E_F$ is the Fermi energy. The site indexes $i$ and $j$ run through atoms within the primitive cell by default, however the inter-site Green's function between any two atoms  of the lattice sites $i'$ and $j'$ could be obtained via
integration over Brillouin zone (BZ):
\begin{equation}
G_{i'j',\sigma}^{mm'} (\varepsilon)= \int_{BZ}  G_{ij,\sigma}^{mm'}(\varepsilon,{\bf k})e^{i{\bf k}(({\bf R}_{i'}-{\bf R}^0_i)-({\bf R}_{j'}-{\bf R}^0_j))} d{\bf k},
\label{scellGf}
\end{equation}
where $G_{ij,\sigma}^{mm'}({\bf k})$ is the inter-site Green's function of the primitive cell for given ${\bf k}$ point, ${\bf R_{i'}}$ -- is the position of atom $i'$ in the lattice, and 
${\bf R^0_i}$ is the position of the same atom within the primitive cell.

The resulting $G_{i'j',\sigma}^{mm'} (\varepsilon)$ is used in the analytic expression
for the exchange integrals as obtained in the Green's function
method~\cite{Liechtenstein1987}:

\begin{equation}
\label{JLEIP}
J_{ij} = -\frac{1}{2\pi}\int\limits_{-\infty}^{E_F}d\varepsilon
\sum_{\substack{mm'\\m''m'''}}
Im(\Delta_{i}^{mm'}G_{ij,\downarrow}^{m'm''}\Delta_{j}^{m''m'''}
G_{ji,\uparrow}^{m'''m}),
\end{equation}
where $G_{ji,\uparrow}^{mm'}$ ($G_{ij,\downarrow}^{mm'}$) is the real-space inter-site Green's function for spin up (down) obtained in Eq.~\eqref{scellGf} and 
\begin{equation}
\Delta_{i}^{mm'} = \int_{BZ} (H^{mm'}_{ii,\uparrow}({\bf k})- 
H^{mm'}_{ii,\downarrow}({\bf k}))d{\bf k}.
\label{delta}
\end{equation}

The proposed scheme allows us to compute per-orbital contribution to the exchange interaction between two atoms. Without spin-orbit coupling the $\Delta_{i}^{mm'}$ matrix is diagonal in the spin subspace,
but it is not necessarily diagonal in the orbital subspace.
However, one may always transform $\Delta_{i}^{mm'}$ to
the diagonal form  (e.g. changing the global coordinate system of the crystal
to the local one, when axes are directed to the ligands; 
or simply diagonalizing on-site Hamiltonian matrix in the 
WF basis set): 
\begin{equation}
\Delta_i^{mm'}=\sum_k T_i^{mk}\tilde \Delta_i^{kk}(T_i^{km'})^*.
\end{equation}
Then Eq.~\eqref{JLEIP} can be rewritten as:
\begin{equation}
\label{LEIP-LCS}
J_{ij}^{kk'} = -\frac{1}{2\pi}\int\limits_{-\infty}^{E_F}d \varepsilon
\sum_{kk'}Im(\tilde \Delta_{i}^{kk} \tilde G_{ij,\downarrow}^{kk'} \tilde \Delta_{j}^{k'k'}
\tilde G_{ji,\uparrow}^{k'k}),
\end{equation}
where 
\begin{equation}
\tilde G_{ij,\sigma}^{kk'}=\sum_{mm'} T_i^{km}\tilde G_{ij,\sigma}^{mm'}(T_j^{m'k'})^*.
\end{equation}
Eq.~\eqref{LEIP-LCS} allows to calculate exchange coupling between the 
$k$th orbital on site $i$ and the $k'$th orbital on site $j$.

In the end of this section we would like to stress that one should carefully chose the orbital set used in the projection procedure. First of all, technically it should be the set and the energy window for the projection, which give the band structure identical (or close to) initial. Secondly, this set should be physically reasonable. E.g. if one deals with compounds (like NiO and KCuF$_3$), where the main exchange mechanism is expected to be superexchange via, e.g. ligand $p$ orbitals, then corresponding states have to be included in the projection procedure. This, in turn, provides an additional tool to study the exchange paths and mechanism of the magnetic coupling, whether it is due to direct or superexchange.

\section{Results and Discussion}

\subsection{NiO}
\begin{figure}[t!]
\centerline{\includegraphics[width=0.7\columnwidth,clip]{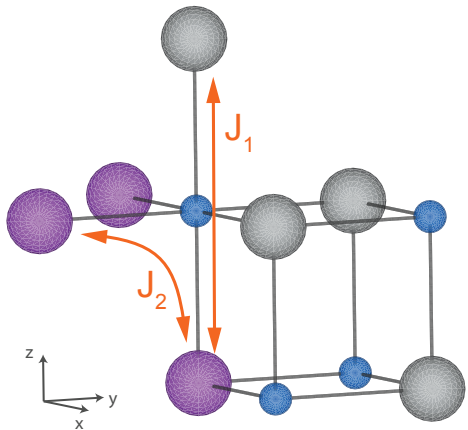}}
\caption{(color online) Schematic view of the NiO crystal structure.
The blue spheres denote oxygen ions, white the gray and magenta spheres denote two 
magnetic types of Ni. The figure was drawn using VESTA~\cite{VESTA} software.}
\label{fig:NiO}
\end{figure}
NiO is one of the typical systems on which different
calculation schemes are tested. It is a charge-transfer 
insulator with a band gap $\sim$ 4 eV~\cite{Huefner-84}
and local magnetic moment of 1.77$\mu_B$\cite{Fender-68}.
NiO crystallizes in the rocksalt (NaCl) structure and exhibits an
antiferromagnetic ordering of type-II fcc (AFM II-type)~\cite{Shull-51}, with planes of 
opposite spins being repeated in alternating order along [111], see 
Fig.~\ref{fig:NiO}. This type of magnetic ordering is 
due to the strong next-nearest-neighbor (nnn) 
coupling between nickel ions via oxygens $2p$ shell. 
The N\'{e}el temperature is T$_N$= 523 K\cite{Tomlinson-55}.

Since accounting for strong electronic correlations
is crucial in the case of NiO~\cite{Anisimov1991a}, we used the 
LSDA+U method~\cite{Anisimov1997} for the calculation of electronic and magnetic
properties. The on-site Coulomb repulsion and intra-atomic 
Hund's rule exchange parameters were chosen to be
$U=8.0$~eV and $J_H=0.9$~eV, respectively~\cite{Anisimov1991a}.
We used the Perdew-Zunger exchange-correlation 
potential~\cite{Perdew1981}, 45 Ry and 360 Ry for the charge density 
and kinetic energy cutoffs, and 512 k-points in the 
Brillouin-zone (BZ).
The unit cell consists of two formula units to simulate AFM II-type.
\begin{figure}[t!]
\resizebox{\columnwidth}{!}{\input{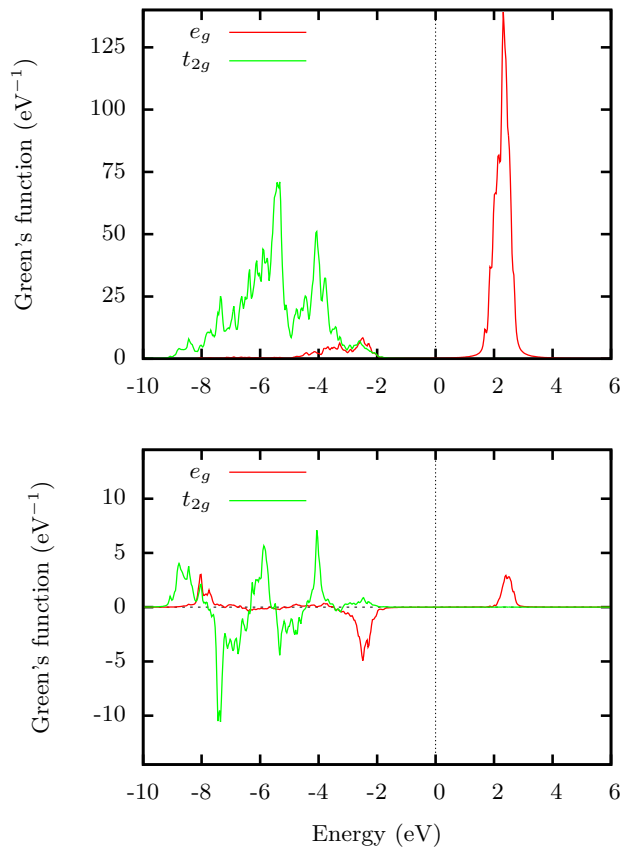}}
\caption{(color online)  Imaginary part of the spin-up on-site (upper panel) Green's function of Ni ion and inter-site (lower panel) Green's function for the pair of the Ni ions along $c$ axis (i.e. corresponding to $J_1$). The Green's function for the $e_g$ states is shown by solid black curve, and for the $t_{2g}$ states by the solid red curve. Zero energy corresponds to the Fermi level.}
\label{fig:NiO_green_inter}
\end{figure}

First of all, we have calculated the dominating exchange interactions for the Heisenberg model \eqref{Heis_q} between
second nearest neighbors, $J_1$ (see Fig.~\ref{fig:NiO}), using
conventional total energy technique and obtained 
$J_1=18.8$ meV, which agrees extremely well with
experimental estimation of $J_1=$19.0 meV~\cite{Hutchings-72}.

The small effective Hamiltonian used for the Green's function 
calculation according to~\eqref{Green} was obtained
by the Wannier function projection procedure as described
in Sec.~\ref{Method}. The Wannier functions were constructed 
as a projection of the Ni $3d$ and O $2p$ pseudoatomic orbitals onto 
subspace of Bloch functions defined by the 16 
energy bands, which predominantly have the Ni $3d$ and O $2p$ 
character: 2 formula units $\times$ (5 Ni $3d$ plus
3 O $2p$ orbitals)=16.

The exchange constants calculated by the Green's function
method are $J_1$ = 18.9~meV, and $J_2$ = -0.4~meV,
and agree with both the total energy and experimental 
estimations. Moreover, they allow to perform an analysis of
partial contributions from different orbitals. An orbital resolved matrix (in meV) for the largest
exchange interaction $J_1$ between the next nearest neighbors along $z$ ($c$)
direction (calculated according to \eqref{LEIP-LCS}) is given as
\begin{equation}
J^{mm'}_1 = 
\begin{pmatrix}
 -18.9 & 0 &  0 &  0 &  0 \\
   0 & 0 &  0 &  0 &  0 \\
   0 & 0 &  0 &  0 &  0 \\
   0 & 0 &  0 &  0 &  0 \\
   0 & 0 &  0 &  0 &  0
\end{pmatrix}.
\end{equation}
Here the following order of the $3d$-orbital is used: $3z^2-r^2$, $zx$, 
$zy$, $x^2-y^2$, $xy$; and the axes of the coordinate
system are shown in Fig.~\ref{fig:NiO}. Thus, one may see that the exchange coupling between the next nearest neighbors is due to overlap between $3z^2-r^2$ orbitals centered on different sites. This is the $180^{\circ}$ superexchange interaction via the $2p_z$ orbital of the oxygen sitting between two Ni ions in the $z$ ($c$) direction, which has to be strong and antiferromagnetic (AFM)
according to Goodenough-Kanamori-Anderson rules~\cite{Goodenough}.
In contrast, the exchange interaction between nearest neighbors, $J_2$,
occurs via two orthogonal $p$ orbitals and is expected to be
weak and ferromagentic (FM)~\cite{Goodenough}.

The imaginary parts of the on-site and inter-site Green's functions are shown in Fig.~\ref{fig:NiO_green_inter}. The inter-site Green's function (lower panel) corresponds to the strongest $180^{\circ}$ exchange
coupling, $J_1$. 
The exchange interaction \eqref{JLEIP} is the energy integral of two Green's functions and two $\Delta$-functions, which do not depend on $\epsilon$. Therefore it is important to explore an energy dependence of the Green's function.

One can see that the on-site Green's function (upper panel) doesn't change its sign over the entire energy interval and after normalization the function is exactly equals to density of electronic states. The energy integral of the on-site Green's function up to the Fermi level gives the total number of electrons on corresponding orbitals. This value is predictable and slight changes to the on-site Green's function peaks positions and widths will not change resulting number of electrons significantly.

The inter-site Green's function, shown in the lower panel of Fig.~\ref{fig:NiO_green_inter} changes its sign several times. It means that in a general case the energy integral up to the Fermi level has an unpredictable sign and the value strongly depends on the Green's function peak positions and widths, i.e. on band structure calculation results.

\subsection{KCuF$_3$}
\begin{figure}[t!]
\centerline{\includegraphics[width=0.7\columnwidth,clip]{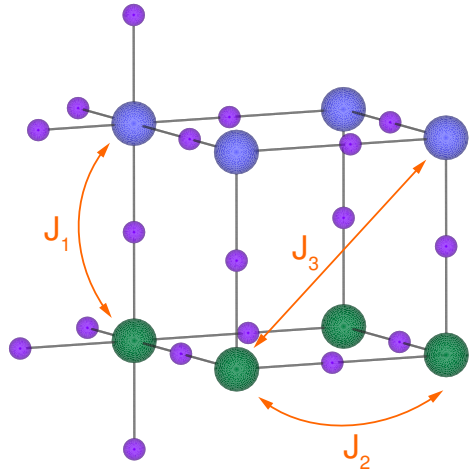}}
\caption{(color online) Schematic view of the KCuF$3$ crystal structure. The blue and green spheres denote Cu ions of two different types, while the violet spheres denote F ions. The potassium ion in the center of the cell is not shown for clarity.}
\label{fig:KCuF3}
\end{figure}

KCuF$_3$ is renowned due to its orbital order, which defines
its magnetic properties. The single hole in the $e_g$ subshell of Cu$^{2+}$ ion (its electronic configuration is $3d^9$) is localized on the
alternating $z^2 - x^2$ and 
$z^2 - y^2$ orbitals present in the $ab$ plane (i.e. antiferro-orbital order),
which results in the weak ferromagnetic coupling
in this plane. In contrast, there is a ferro-orbital 
ordering in the $c$ direction, which leads to a strong
antiferromagnetic interaction along this axis. As a
result in the essentially three dimensional (3D) crystal
one may observe the formation of nearly ideal
one-dimensional antiferromagnetic Heisenberg 
chains~\cite{KK-UFN,KK-JETP}. 
\begin{figure}[t!]
\resizebox{\columnwidth}{!}{\input{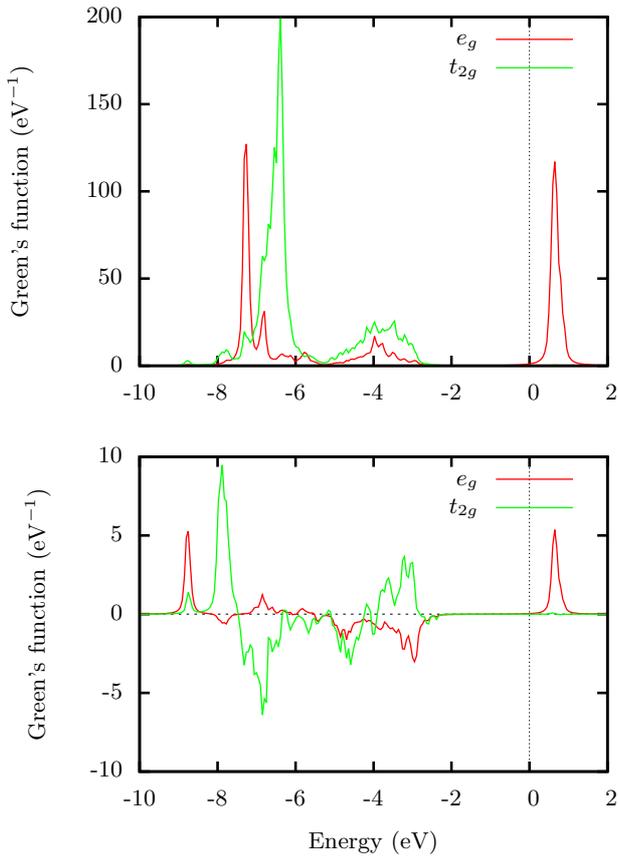}}
\caption{(color online)  Imaginary part of the spin-down on-site (upper panel) Green's function of Cu ion and inter-site (lower panel) Green's function for two Cu ions corresponding to $J_1$. The Green's function for $e_g$ states is shown by solid black line, for $t_{2g}$ states by the solid red one. Zero energy corresponds to the Fermi level.}
\label{fig:KCuF3_2G}
\end{figure}

The compound has a distorted cubic perovskite crystal structure (shown in Fig.~\ref{fig:KCuF3}) with space group $I4/mcm$. The copper ions have octahedral fluorine surrounding. These octahedra are elongated along one of the
directions. At room temperature, there are two different structural polytypes with antiferro ($a$-type) and ferro-like ($d$-type) stacking of the $ab$ planes along the $c$ axis~\cite{Okazaki1969}.

Altogether, the electronic and structural properties of KCuF$_3$ have previously been intensively studied by employing density functional theory and its extensions like the DFT+U approach~\cite{Anisimov1991}. The DFT+U calculations led to a correct insulating ground state with the spin and orbital ordering~\cite{Liechtenstein1995,Medvedeva2002,Binggeli2004} that are in agreement with experimental data. We used the GGA+U approach as a starting point for the exchange interaction parameters calculation.

For the density functional calculations, we used the Perdew-Burke-Ernzerhof~\cite{Perdew1996} GGA exchange-correlation functional together with Vanderbilt utrasoft pseudopotentials. We set the kinetic energy cutoff to  50 Ry (400 Ry) for the plane-wave expansion of the electronic states (core-augmentation charge). The self-consistent calculation was performed with the 4$\times$4$\times$4 Monkhorst-Pack k-point grid. We set the effective on-site Coulomb interaction as $U_{eff} = U - J_H =  6.6$~eV~\cite{Binggeli2004}. To reproduce the magnetic and orbital ordering of the polytype {\em a}, we used a cell containing four formula units.

The basis of the WFs has a dimension of 56. It includes 
20 Cu-$d$ like WFs (5 functions for every Cu site) and 
36 F-$p$ like WF.
We generated the Cu WF using a linear combination of pseudoatomic Cu-$d$ orbitals to obtain a more clear physical basis for the Green's function formalism. 

The strongest exchange interaction was found to be between nearest Cu ions along the $c$ axis, $J_1$ = 17~meV (antiferromagnetic). As it was mentioned above this is because of the ferro-orbital order in this
direction, given by $J_1 \sim t^2/U$ (where $t$ is corresponding hopping integral). The calculated value agrees with different experimental estimations of $J_1$, which was found be 16.1~meV~\cite{Iio1978} using analysis of the specific heat data,  16.2~meV~\cite{Kadota1967} based on the temperature dependence of the magnetic susceptibility and 17~meV\cite{Hutchings1979} or 17.5~meV~\cite{Satija1980} in neutron measurements.

The exchange coupling in the $ab$ plane, given by $J_{2} \sim t^2 J_H/ U^2$, has to be much
weaker, since there is an
antiferro-orbital order. Our calculations give $J_{2}$ = 0.5~meV. The additional ``diagonal'' exchange, $J_3$
was estimated to be -1~meV. 

The on-site and inter-site Green's functions for KCuF$_3$ are shown in Fig.~\ref{fig:KCuF3_2G}. 
The main contribution to exchange interaction in $c$ direction comes 
from the overlap between the similar WFs centered on different Cu ions 
(i.e. $z^2-y^2$/$z^2-y^2$ or $z^2-x^2$/$z^2-x^2$).

\subsection{FeO}
FeO together with NiO is one of the most studied monoxides. The crystal structure of these oxides is quite similar and shown in Fig.~\ref{fig:NiO} (there are small rhombohedral distortions in the magnetically ordered phase of FeO), but magnetic properties of FeO strongly depend on the amount of defects in samples. The ordered moment changes from 3.2 to 4.5 $\mu_B$, while N\'{e}el temperature $T_N$ is $\sim$200~K (FeO orders in the AFM II-type structure; the same as NiO)~\cite{Fjellvag1996}. Due to geophysical importance of FeO the investigations were mostly concentrated on the pressure dependence of its magnetic properties. Possible presence of the pressure driven spin-state transition was studied by different methods starting from the conventional DFT calculations to more elaborated methods based on the dynamical mean-field theory (DMFT)~\cite{Isaak1993,Cohen1997,Shorikov2010}. However, in addition to this transition there is also unconventional change of $T_N$ with the pressure~\cite{Struzhkin1999}. Thorough study of this effect in a wide pressure range is beyond the scope of the present paper, but we estimated the change of the N\'{e}el temperature for moderate pressures.

We used experimental crystal structure for zero pressure~\cite{Fjellvag1996}, and optimized it (keeping the symmetry) for the pressure of 15 GPa. Standard PBE pseudopotentials from the Quantum-ESPRESSO pseudopotentials library were used for the self-consistent ground state calculation. The plane-wave energy cutoff value was set to 45~Ry. Integration over the reciprocal cell was performed on 16x16x16 regular k-points grid. The Hubbard's parameters  $U$=5 eV and $J_H$=0.9 eV were calculated by one of us for FeO in the same pseudopotential code previously\cite{Shorikov2010}.
The WF basis consists of 16 Wannier functions. It includes states with Fe-d and O-p orbitals symmetry for two formula units.

The second nearest neighbor exchange coupling (see Fig. \ref{fig:NiO}) was found to be  $J_1$=2.1 meV for the Heisenberg model written in Eq.\eqref{Heis_q}. In the mean-field approximation the N\'{e}el temperature for the fcc lattice and AFM of II-type can be estimated as 6$J_1 \frac 13 S(S+1)$, which gives $T_N \sim 300$ K, while experimental $T_N^{exp} \sim 200$ K.
This is a common feature of the mean-field theories to overestimate the transition temperature in 1.5-2 times (e.g., the situation in NiO is rather similar; if one would even use experimental $J_1=19$ meV, the N\'{e}el temperature will be strongly overestimated). What is more representative is the ratio between $T_N$ for different pressures. Experimentally $T_N^{P=15}/T_N^{P=0} \approx 1.45$,\cite{Struzhkin1999}
while theoretically we obtained $T_N^{P=15}/T_N^{P=0}=1.4$. 
Thus, one doesn't need to use such a sophisticated techniques as DMFT to describe pressure dependence of the N\'{e}el temperature in FeO (at least for moderate pressures), which can be explained by the modification of average Fe-O-Fe distance. Indeed, in the Mott-Hubbard systems the superexchange between half-filled orbitals is defined by effective hopping parameter $\tilde{t}_{dd}$ via ligand $2p$ orbitals\cite{Goodenough}:
\begin{eqnarray}
\label{t2U}
J \sim \frac {\tilde{t}_{dd}^2}{U}
\end{eqnarray}
and $\tilde{t}_{dd} \sim \frac {t_{pd}^2}{\Delta_{CT}}$, where $\Delta_{CT}$ is the charge-transfer energy~\cite{ZSA} and $t_{pd}$ is the hopping between ligand $p$ and metal $d$ orbitals. Since this hopping scales as $t_{pd} \sim 1/d^{7/2}$~\cite{Harrison1999}, where $d$ is the distance between ligand and transition metal ion, then
\begin{eqnarray}
\label{TN-d}
\frac{T_N^{P=15}} {T_N^{P=0}} = \Big(\frac{d^{P=0}_{Fe-O}}{d^{P=15}_{Fe-O}}\Big)^{14}.
\end{eqnarray}
Such a crude estimation surprisingly works quite well. According to our GGA calculations going from zero to 15 GPa pressure $d_{Fe-O}$ changes on 2.7\%. Then according to \eqref{TN-d}
$\frac{T_N^{P=15}} {T_N^{P=0}} = 1.45$, exactly as 
observed experimentally~\cite{Struzhkin1999}.

\subsection{Li$_2$MnO$_3$}  
\begin{figure}[t!]
\centerline{\includegraphics[width=0.8\columnwidth,clip]{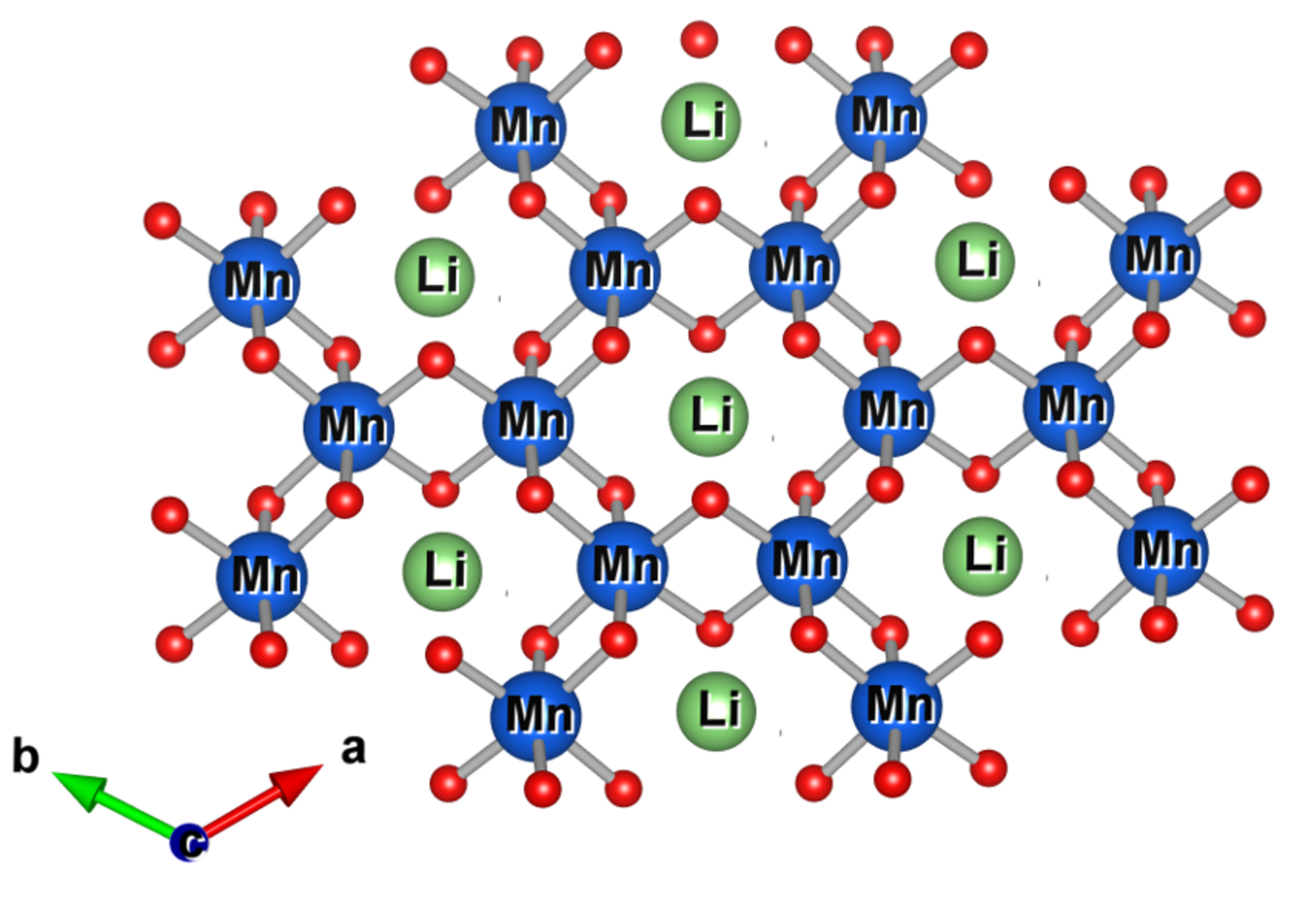}}
\caption{(color online) Crystal structure of Li$_2$MnO$_3$. Mn ions shown by blue balls form are in octahedral surrounding of the O ions (red balls)  and form  honeycomb lattice, with Li (green balls) in the center of the honeycombs. These 2D hexagonal planes are stacked in the $c$ direction with Li ions in between.}
\label{fig:Li2MnO3}
\end{figure}

Compounds with general formula A$_2$BO$_3$, where $A$ is an alkali metal, Li or Na, and $B$ is a metal have layered crystal structure with B ions forming honeycomb lattice, see Fig.~\ref{fig:Li2MnO3}. They attract much attention not only due to possible technological application as battery cathode materials~\cite{Todorova2011}, but also represent special interest for the fundamental science. E.g. Na$_2$IrO$_3$ is considered as a possible realization of the Kitaev model~\cite{Jackeli2009}, while Li$_2$RuO$_3$ shows unusual valence bond liquid phase at high temperatures~\cite{Kimber2013} and spin gapped state below 540 K (at least in polycrystalline samples)~\cite{Miura2007,Wang2014}. In contrast to these systems in Li$_2$MnO$_3$ the long range antiferromagnetic state is formed at T$_N$=36 K with all Mn neighbors in the $ab$ plane ordered AFM~\cite{Lee2012}. This result is rather unexpected, since in the 90$^{\circ}$ Mn-O-Mn geometry one might expect strong FM interaction between half-filled $t_{2g}$ and empty $e_g$ orbitals of Mn$^{4+}$ ions~\cite{Lee2012,Goodenough,Streltsov2008}. 

We performed GGA and GGA+U calculations of the exchange parameters in Li$_2$MnO$_3$ using Perdew-Burke-Ernzerhof~\cite{Perdew1996} exchange-correlation potential. The crystal structure was taken from Ref.~\cite{Lee2012} for T=6 K. The magnetic structure is AFM G-type, when all neighboring Mn are AFM coupled~\cite{Lee2012}. The kinetic energy cutoff was chosen to be 45 Ry (450 Ry) for the plane-wave expansion of the electronic states (core-augmentation charge) and we used  64 k-points for the integration over the BZ.  

\begin{figure}[t!]
\resizebox{\columnwidth}{!}{\input{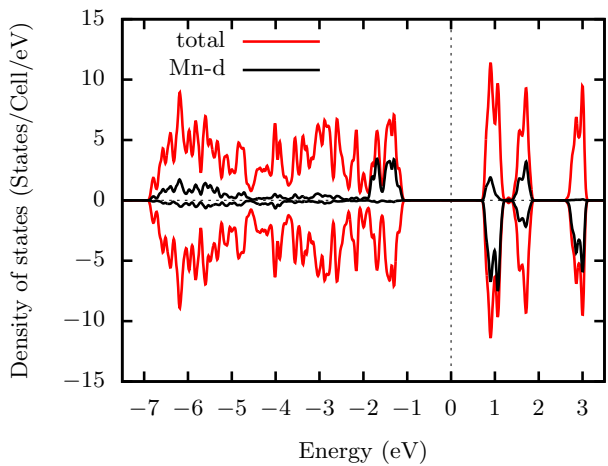}}
\caption{(color online)  Spin polarized density of electronic states for Li$_2$MnO$_3$ obtained in the magnetic GGA calculation. Zero energy corresponds to the Fermi level.}
\label{fig:Li2MnO3_DOS}
\end{figure}

The magnetic moments on Mn ions in the GGA approach were found to be 2.5 $\mu_B$, which is consistent with 4+ oxidation state. The total and partial DOS are shown in Fig.~\ref{fig:Li2MnO3_DOS}.
This is the feature of the Mn$^{4+}$ ion with the half-filled $t_{2g}$ sub shell (electronic configuration $3d^3$), that the spin splitting (i.e. the splitting between spin majority and spin minority sub bands) is quite large and therefore already magnetic GGA calculation gives insulating ground state with the band gap 1.9 eV. On the one hand, this is much larger than experimental activation energy $\Delta \sim 0.7$ eV deduced from the resistivity measurements\cite{Lee2012}, which, however, cannot be considered as a direct and precision way of the estimation of the band gap. On the other hand, this strongly suggests that the Hubbard correction, $U$, is not that important for the descriptions of the top of the valence and the bottom of the conduction bands. Indeed, many other Mn oxides can be described by the LSDA or GGA methods without account of any Hubbard correlations~\cite{Solovyev1996,Weht2001,Park2010,Whangbo2007}. 

Our GGA+U calculation shows that even quite large $U=4.5$ eV only slightly increases the value of the band gap (on 0.3 eV), which shows that the band gap is indeed defined by the spin splitting (as clearly seen from Fig.~\ref{fig:Li2MnO3_DOS}) and not by the Coulomb correlations. Therefore the use of the GGA approximation seems to be plausible for the description of the magnetic properties of Li$_2$MnO$_3$. 
This additionally allows us to test the Green's function approach for the calculation of the exchange constants without Hubbard's $U$.  

We found that in the GGA approximation exchange coupling between nearest neighbors is $J=23$~K (AFM) for the Heisenberg model defined in Eq.~\eqref{Heis_q}. In the mean-field approximation this gives Curie-Weiss temperature $\theta_{GGA}=87$~K. This is again somewhat larger than experimental $\theta_{exp} \sim 50-60$~K~\cite{Lee2012}, but it agrees with what one may expect from the mean-field theory. An account of the on-site Coulomb repulsion in the GGA+U calculation leads to gradual growth of the FM component and results in total exchange $J=-16$ K (FM) for $U=4.5$ eV and $J_H=0.9$ eV (as were used, e.g., in NaMn$_7$O$_{12}$\cite{Streltsov2014d} or in Mn$_4$(hmp)$_6$\cite{Streltsov2014c}), which agrees with Goodenough-Kanamori-Anderson rules~\cite{Lee2012,Goodenough}, but is inconsistent with experiment~\cite{Lee2012}. 

Thus, the results of the GGA calculations, where Li$_2$MnO$_3$ turns out to be a Slater insulator with the band gap appearing due to a spin splitting, seem to be reasonable. 
In the first order of the perturbation theory the exchange interaction in this situation is expected to be AFM. It can be described not by Eq.~\eqref{t2U}, but rather as
\begin{eqnarray}
J \sim \frac{t_{dd}^2}{\Delta_{exc}},
\end{eqnarray}
where $\Delta_{exc}$ is the exchange splitting, which in the GGA is given by the sublattice magnetization $M$ and Stoner parameter $I$ as $\Delta_{exc} = IM$. 

\section{Conclusion}
We have presented the implementation of the Green's function approach for the Heisenberg model exchange parameters calculation. 
The localized electronic states were described by the Wannier functions with the symmetry of atomic orbitals. This basis set allowed us to overcome the limitations of modern plane-wave based calculation 
schemes and perform a complex analysis of the inter-site exchange interaction with the density functional theory or its extensions 
such as DFT+U. The results were tested on four transition metal compounds: NiO, FeO, KCuF$_3$, and Li$_2$MnO$_3$. The obtained values are in a good agreement with experimental estimations.

\section{Acknowledgments}
We thank A.\,V.~Lukoyanov, and A.~Pitman for valuable comments and J.-G. Park for the communications about layer A$_2$BO$_3$ compounds.
The present work was supported by the grant of the Russian Scientific Foundation (project no. 14-22-00004).

\bibliography{main}

\end{document}